\documentclass[12pt]{article}

\usepackage{amssymb}
\usepackage{amsmath}
\usepackage{amscd}
\usepackage{latexsym}
\usepackage{graphicx}

\usepackage{cite}

\newcommand{\be}{\begin{equation}}
\newcommand{\ee}{\end{equation}}
\newcommand{\Dlt}{\Delta}

\newcommand{\bS}{{\bf S}}
\newcommand{\bB}{{\bf B}}
\newcommand{\bt}{\beta}

\newcommand{\al}{\alpha}
\newcommand{\ra}{\rightarrow}

\newcommand{\gm}{\gamma}
\newcommand{\om}{\omega}

\newcommand{\rgl}{\rangle}
\newcommand{\lgl}{\langle}

\begin{document}

\begin{center}

{\Large{\bf Triggering spin reversal in nanomolecules and nanoclusters on 
demand} \\ [5mm]

V.I. Yukalov$^{1,2}$ and E.P. Yukalova$^{3}$ }  \\ [3mm]

{\it
$^1$Bogolubov Laboratory of Theoretical Physics, \\
Joint Institute for Nuclear Research, Dubna 141980, Russia \\ [2mm]

$^2$Instituto de Fisica de S\~ao Carlos, Universidade de S\~ao Paulo, \\
CP 369, S\~ao Carlos 13560-970, S\~ao Paulo, Brazil \\ [2mm]

$^3$Laboratory of Information Technologies, \\
Joint Institute for Nuclear Research, Dubna 141980, Russia } \\ [3mm]

{\bf E-mails}: {\it yukalov@theor.jinr.ru}, ~~ {\it yukalova@theor.jinr.ru}

\end{center}

\vskip 1cm

\begin{abstract}
Spin reversal in magnetic nanomolecules and nanoclusters is considered. 
A method is suggested allowing, from one side, to keep for long time 
magnetic polarization in a metastable state and, from the other side, for 
starting the reversal process at any required time. This method can find 
applications for the operation of storage memory devices and for the regulation 
of processes in spintronics.  
\end{abstract}

\vskip 2cm
{\parindent=0pt
{\bf Keywords}: magnetic nanomolecules, spin reversal, data storage, memory 
devices, triggering condition  }

\newpage

\section{Introduction}

Magnetic nanomolecules 
\cite{Barbara_1,Caneschi_2,Friedman_3,Miller_4,Craig_5,Liddle_6,Rana_7} and 
nanoclusters \cite{Kodama_8,Kudr_9} of different nature are of extraordinary 
technological importance with a myriad of applications that permeate daily life. 
Some properties exhibited by quantum dots, whose sizes are in the nanometer 
range, are similar to those of nanomolecules and nanoclusters, because of which 
quantum dots are often called ``artificial atoms" \cite{Birman_10}. There also 
exist magnetic quantum dots \cite{Schwartz_11,Koole_12,Mahajan_13,Tufani_14} 
enjoying some of the properties of magnetic nanomolecules and nanoclusters.

Most applications of nanomagnets involve rotating the magnetization orientation 
of one or more nanomagnets. A single-domain nanomagnet is a nanomagnet of 
sufficiently small size having a single ferromagnetic domain in which all the 
spins point in the same direction because of strong exchange interaction between 
them. That direction is the direction of its magnetization. If we rotate the 
magnetization, then all the spins in the single-domain nanomagnet rotate together 
in unison, so the magnetization acts like a giant single spin. This is called 
coherent rotation. If the process of coherent rotation is sufficiently fast, which 
can be achieved by connecting the sample to a resonance electric circuit 
\cite{Krishnan_15}, magnetic nanomolecules and nanoclusters can emit coherent 
radiation \cite{Yukalov_16,Yukalov_17,Yukalov_18,Yukalov_19,Benedict_20,
Yukalov_21,Yukalov_22}. Sufficiently fast reversal of magnetization, in about 
$10^{-8}$ s, can be realized by acting on the sample with alternating magnetic 
fields \cite{Shuty_23}. Short intensive laser pulse can reverse magnetization 
in about $10^{-9}$ s \cite{Deb_24}. We consider spin reversal caused by the 
action of resonator feedback field 
\cite{Yukalov_16,Yukalov_17,Yukalov_18,Yukalov_19,Yukalov_21,Yukalov_22}, 
which seems to provide the most convenient and, probably, the fastest method 
for spin reversal \cite{Yukalov_25}.  

Among the most important applications of magnetic nanomolecules and nanoclusters, 
it is necessary to mention the possible use of their spins for data storage in 
memory devices, for spin-based quantum computers \cite{Cerletti_26,Morton_27}, 
and for information processing \cite{Yukalov_28}. To serve as an efficient 
storage memory device, the system has to satisfy two main properties contradicting 
each other: 

(i) It can keep information for long time, which is connected with the 
feasibility of keeping intact the fixed spin position. This can be provided by 
a strong anisotropy typical of molecular magnets and clusters.

(ii) The memory can be either quickly erased or changed by reversing spin 
polarization. This means that the magnetic anisotropy could be suppressed 
at any required moment of time.

In this article, we suggest a method allowing for triggering spin reversal in 
magnetic nanomolecules and nanoclusters with large magnetic anisotropy at any 
desired time. Some trapped spin-$1$ atoms in optical latices, e.g. $^{87}$Rb 
atoms in the $F=1$, $m_F=-1$ hyperfine state, can also possess magnetic 
anisotropy \cite{Chung_29}, hence clusters of trapped atoms can also be kept 
in mind.

\section{Main equations}

The Hamiltonian typical of a magnetic nanomolecule or a nanocluster is
\be
\label{1}
\hat H = - \mu_S \bB \cdot \bS + \hat H_A \; ,
\ee
where in the Zeeman term $\mu_S = -g_S \mu_B$ is the magnetic moment of the 
molecule, $g_S$ the molecule Land\'e factor, and $\mu_B$, Bohr magneton. The 
second term is the Hamiltonian describing magnetic anisotropy,
\be
\label{2}
\hat H_A = - DS_z^2 + E( S_x^2 - S_y^2 )  \;  ,
\ee
with the parameters
$$
D = \frac{1}{2} \; ( D_{xx} + D_{yy} ) - D_{zz} \; , \qquad
E = \frac{1}{2} \; ( D_{xx} - D_{yy} ) \; ,
$$
expressed through the tensor of magnetic anisotropy $D_{\al\bt}$.

The sample is inserted into a magnetic coil of a resonance electric circuit 
creating a feedback field $H$. The total magnetic field, acting on the sample, 
\be
\label{3}
 \bB = ( B_0 + \Dlt B ) {\bf e}_z + H {\bf e}_x  + 
B_1 {\bf e}_y \; ,
\ee
consists of a constant field $B_0$ along the axis $z$, an additional regulated 
field $\Delta B$ that can be varied, the feedback field of the resonator coil, 
$H$ in the $x$ direction, and a small anisotropy field $H_1$ along the axis $y$.  

The feedback field $H$ satisfies the equation
\be
\label{4}
  \frac{dH}{dt} + 2\gm H + \om^2 \int_0^t H(t') \; dt' = - 
4\pi\eta_f \; \frac{dm_x}{dt} 
\ee
following \cite{Yukalov_16,Yukalov_17,Yukalov_18,Yukalov_19,Yukalov_25} from the 
Kirchhoff equation. Here $\gamma$ is the resonator attenuation, $\omega$, resonator 
natural frequency, $\eta_f = V/V_c$ is a filling factor, $V$ sample volume, $V_c$, 
coil volume, and the right-hand side is the electromotive force produced by moving 
spins with the magnetization density  
\be
\label{5}
 m_x = \frac{\mu_S}{V} \; \lgl \; S_x \; \rgl \;  .
\ee

For what follows, let us introduce the notations for the Zeeman frequency
\be
\label{6}
  \om_0 \equiv - \; \frac{\mu_S}{\hbar}\; B_0 \; ,
\ee
regulated frequency
\be
\label{7}
 \Dlt \om \equiv - \; \frac{\mu_S}{\hbar}\; \Dlt B \;  ,
\ee
and the transverse frequency
\be
\label{8}
  \om_1 \equiv - \; \frac{\mu_S}{\hbar}\; B_1 \;  .
\ee

The right-hand side of the feedback equation (\ref{4}), which is the electromotive 
force, is proportional to the rate  
\be
\label{9}
\gm_0 \equiv \pi\eta_f \; \frac{\mu_S^2 S}{\hbar V} = 
\pi \; \frac{\mu_S^2 S}{\hbar V_c}   
\ee
that can be called feedback rate. It is useful to define the dimensionless 
feedback field
\be
\label{10}
 h \equiv - \; \frac{\mu_S H}{\hbar \gm_0} \;  .
\ee

Using the commutators
$$
[\; S_x, \; S_y^2 \; ] = - [\; S_x, \; S_z^2 \; ] = 
i (S_y S_z + S_z S_y ) \; ,
$$
$$
[\; S_y, \; S_z^2 \; ] = - [\; S_y, \; S_x^2 \; ] = 
i (S_x S_z + S_z S_x ) \; ,
$$
$$
[\; S_z, \; S_x^2 \; ] = - [\; S_z, \; S_y^2 \; ] = 
i (S_x S_z + S_z S_x ) \;   ,
$$
we derive the Heisenberg equations of spin motion for the $x$-component,
\be
\label{11}
\frac{dS_x}{dt} = - ( \om_0 + \Dlt\om) S_y + \om_1 S_z +
\frac{D+E}{\hbar}\;  (S_y S_z + S_z S_y ) \;  ,
\ee
for the $y$-component,
\be
\label{12}
\frac{dS_y}{dt} =  ( \om_0 + \Dlt\om) S_x - \gm_0 h S_z - \;
\frac{D+E}{\hbar}\;  (S_x S_z + S_z S_x ) \;   ,
\ee
and for the $z$-component,
\be
\label{13}
 \frac{dS_z}{dt} = \gm_0 h S_y - \om_1 S_x + 
\frac{2E}{\hbar} \; (S_x S_y + S_y S_x ) \; .
\ee

It is convenient to introduce the variables
\be
\label{14}
x \equiv \frac{\lgl S_x\rgl}{S} \; ,  \qquad 
y \equiv \frac{\lgl S_y\rgl}{S} \; ,  \qquad 
z \equiv \frac{\lgl S_z\rgl}{S} \; , 
\ee
varying in the interval $[-1,1]$.

We use the decoupling of pair correlations in the form \cite{Yukalov_16,Yukalov_17}
\be
\label{15}
 \lgl \; S_\al S_\bt + S_\bt S_\al \; \rgl = 
\left( 2 \; - \; \frac{1}{S}\right) 
\lgl \; S_\al \; \rgl \lgl \; S_\bt \; \rgl \qquad ( \al \neq \bt) \; ,
\ee
which is exact for spin one-half, since
$$
 S_\al S_\bt + S_\bt S_\al = 0 \qquad 
\left( \al \neq \bt , ~ S = \frac{1}{2} \right) \; ,
$$
and for large spins $S \ra \infty$ that behave classically. 

The feedback-field equation (\ref{4}) becomes
\be
\label{16}
 \frac{dh}{dt} + 2\gm h + \om^2 \int_0^t h(t') \; dt' = 
4\; \frac{dx}{dt} \;  .
\ee

The dimensionless anisotropy parameter 
\be
\label{17}
A \equiv \frac{\om_D + \om_E}{\om_0}
\ee
is expressed through the effective anisotropy frequencies
\be
\label{18}
 \om_D \equiv ( 2 S - 1) \;\frac{D}{\hbar} \; , \qquad
\om_E \equiv ( 2 S - 1) \;\frac{E}{\hbar} \;  .
\ee
Then equations (\ref{11}), (\ref{12}) and (\ref{13}) read as
$$
\frac{dx}{dt} = -\om_0 ( 1 + b - Az) y + \om_1 z \; , \qquad
\frac{dy}{dt} = \om_0 ( 1 + b - Az) x - \gm_0 h z \; ,
$$
\be
\label{19}
 \frac{dz}{dt} = 2\om_E x y -  \om_1 x + \gm_0 h y \; ,  
\ee
where 
\be
\label{20}
 b \equiv \frac{\Dlt\om}{\om_0} = -\; 
\frac{\mu_S\Dlt B}{\hbar\om_0} \;  .
\ee

The feedback-field equation (\ref{16}) can be rewritten as the differential 
equation
\be
\label{21}
 \frac{d^2h}{dt^2} + 2\gm\; \frac{dh}{dt} + \om^2 h = 
4 \; \frac{d^2x}{dt^2} \;  .
\ee
The initial conditions are
$$
x_0 = x(0) \; , \qquad y_0 = y(0) \; , \qquad z_0 = z(0) \; ,
$$
\be
\label{22}
h_0 = h(0) =0 \; , \qquad \dot{h_0} = \dot{h}(0)= 0 \; ,
\ee
where the overdot means time derivative.

\section{Triggering condition}

Let the system be prepared so that the spin be polarized along the axis $z$ with 
the initial value $z_0>0$, while $x_0=y_0=0$. This direction of spin, formed by 
the spins of electrons, under the magnetic field $B_0 > 0$, is metastable and 
the spin tends to reverse to the direction down. As we know 
\cite{Yukalov_16,Yukalov_17,Yukalov_18,Yukalov_19}, if the anisotropy parameter 
$A$ were small, the process of spin reversal would start from the very beginning. 
However, in nanomolecules ad nanoclusters, this parameter $A$ can be of order 
one and larger. Then the spin is blocked and can stay intact during quite long 
time of the order of the longitudinal relaxation time $T_1$ due to spin-phonon 
interactions. This time can be estimated by the Arrhenius law
$$
T_1 = \tau_0\exp\left( \frac{U_{eff}}{k_B T} \right) \;  ,
$$
with $U_{eff}$ being the effective barrier separating the directions of spin 
up and spin down. Depending on temperature $T$, the relaxation time $T_1$ can 
be of the order of hours an even months. 

To start spin reversal, it is required that the resonance be realized, when 
the effective Zeeman frequency of spin rotation would be close to the resonator 
natural frequency $\om$. In our case, as equations (\ref{19}) show, the effective 
Zeeman frequency is
\be
\label{23}
 \om_{eff} = \om_0 ( 1 + b - Az) \; .
\ee
Since the spin polarization $z$ changes from $z=1$ to $z=-1$, the effective 
frequency $\omega_{eff}$ varies by a large quantity $2A \omega_0$, hence no 
permanent resonance can be realized. 

Suppose the sample is in a metastable spin state $z_0 > 0$, as described above, 
and can keep that state for very long times due to large magnetic anisotropy. Let 
$\omega = \omega_0$. The regulated field $b = b(t)$ is a function of time. We put 
forward the suggestion that the reversal process can be triggered at any chosen 
time $\tau$ provided the resonance is realized at this time, so that 
\be 
\label{24}
  \om_{eff} (\tau) = \om \; .
\ee
This implies that at the triggering time $\tau$ there should be 
\be
\label{25}
 b(\tau) = Az_0 \qquad ( \om = \om_0 ) \;  .
\ee
Taking the initial spin polarization as $z_0 = 1$, we get the triggering 
condition
\begin{eqnarray}
\label{26}
b(t) =\left\{ \begin{array}{ll}
0 , ~ & ~ t < \tau \\
A , ~ & ~ t \geq \tau 
\end{array} \; . \right.
\end{eqnarray}

\section{Numerical solution}

To study spin dynamics, we need to solve the system of equations (\ref{19}) 
and (\ref{21}). Before the solution, let us estimate the typical parameters 
entering the equations.

One of the well studied magnetic nanomolecules is the dodecanuclear manganese 
cluster with the chemical formula 
[Mn$_{12}$O$_{12}$(CH$_3$COO)$_{16}$(H$_2$ O)$_4$]$\cdot 2$CH$_3$COOH$\cdot$4H$_2$O, 
which is briefly named Mn$_{12}$. It has spin $S=10$ and blocking temperature 
$3.3$ K. Spin polarization can be kept frozen for very long times depending on 
temperature. The time $T_1$ is given by the Arrhenius law, where 
$\tau_0\sim (10^{-7}-10^{-8})$ s. Thus at $T=3$ K, the spin is frozen for one hour 
and at $2$ K, for $2$ months. The molecule radius is around $10^{-7}$ cm and the 
volume of a molecule is $V\sim 10^{-20}$ cm$^3$. The Zeeman frequency depends on the 
magnetic field $B_0$. Thus for $B_0=1 T=10^4$ G, we have $\om_0\sim 10^{11}$ 
s$^{-1}$. The feedback rate (\ref{9}) is $\gm_0\sim 10^8$ s$^{-1}$. The anisotropy 
parameter $D \sim 10^{-16}$ erg, while $E$ is very small, being close to zero. Then 
$D/k_B\approx 0.6$ K and $D/\hbar\sim 10^{11}$ s$^{-1}$. The anisotropy frequency 
(\ref{18}) is $\om_D\sim 10^{12}$ s$^{-1}$. Then the anisotropy parameter (\ref{17}) 
is $A\sim 10$.

The other often mentioned nanomolecule is the octanuclear iron cluster, with the 
formula [Fe$_8$O$_2$(OH)$_{12}$(tacn)$_6$]$^{8+}$, where ``tacn" stands for organic 
ligand triazacyclononane. The abbreviation for this molecule is Fe$_8$. Its properties 
are close to Mn$_{12}$. Spin is $S=10$. Blocking temperature is $T_B\approx 1$ K. The 
molecule volume $V\sim 10^{-20}$ cm$^3$. The anisotropy parameters are 
$D\sim 0.4\times10^{-16}$ erg and much smaller $E$. Thus $D/k_B = 0.27.5$K and 
$E/k_B=0.046$ K, or $D/\hbar\sim 4\times 10^{10}$ s$^{-1}$ and $E/\hbar\sim 10^{10}$ 
s$^{-1}$. The anisotropy frequencies (\ref{18}) are $\om_D \sim 4 \times 10^{11}$ 
s$^{-1}$ and $\om_E\sim 10^{10}$ s$^{-1}$. Then the anisotropy parameter (\ref{17}) 
is $A \sim 1 - 4$.   

Typical parameters of Co, Fe, and Ni nanoclusters are as follows. The volume 
is close to $V\sim 10^{-20}$ cm$^3$, the number of atoms in a cluster is 
$N\sim 10^3-10^4$, the total cluster spin is $S\sim 10^3-10^4$, blocking 
temperature is $T_B\sim (10 - 100)$ K. For the magnetic field $B_0 \sim 1$ T, 
the Zeeman frequency is $\om\sim 10^{11}$ s$^{-1}$. The feedback rate (\ref{9})
is $\gm_0\sim (10^{10}-10^{11})$ s$^{-1}$. The anisotropy parameters are 
$D\sim 10^{-20}$ erg and $E\sim 10^{-19}$ erg, which gives $D/\hbar\sim 10^7$ 
s$^{-1}$ and $E/\hbar\sim 10^6$ s$^{-1}$. Therefore $\om_D\sim (10^{10}-10^{11})$ 
s$^{-1}$ and $\om_E\sim (10^{9}-10^{10})$ s$^{-1}$. Thus the dimensionless 
anisotropy parameter (\ref{17}) is $A\sim 0.1-1$.

Numerical solution of equations (\ref{19}) and (\ref{21}) is accomplished for 
the initial conditions $x_0=y_0=0$, $z_0=1$, $h_0=0$, and $\dot{h}=0$. Frequency 
parameters are measured in units of $\gm_0$ and time, in units of $\gm_0^{-1}$. 
In the figures, the accepted parameters are: $\om=\om_0=10$, $\om_E=\om_1= 0.01$,
$\gm=1$, and $A=1$. 

In order to emphasize the importance of switching on the regulated field $b$ 
exactly to the anisotropy parameter $A$, in Fig. 1, we relax this requirement, 
by switching on the regulated field $b$ to different values $B=b(\tau)$ and show 
that the reversal starts exactly at the given delay $\tau$ only when $B=A=1$. 
When $B$ does not equal $A$, the beginning of the reversal shifts to times larger 
than $\tau$. If the regulated field is not switched on at all, hence $b = B = 0$, 
there is no spin reversal. 

In Fig. 2, the triggering condition (\ref{26}) is preserved, with $B=A=1$, while 
the delay time $\tau$ is varied. As is seen, the spin reversal starts exactly at 
the given delay time $\tau$. The resonance condition (\ref{24}) is valid at the 
time $\tau$, but later on, when $t>\tau$, the resonance condition (\ref{24}) is 
not satisfied. This explains the appearance of the tails with the slower spin 
reversal.   
 
Figure 3 shows the behavior of the transverse spin component $x$, when the 
triggering resonance, with $B=A=1$, is switched on at the time $\tau=10$. In Fig. 3a, 
the interval of time is $[0,40]$, while in Fig. 3b, the interval is $[10,20]$. 
Figure 4 displays the behavior of the feedback field $h$ for the same parameters 
as in Fig. 3.

\section{Conclusion}

For the operation of memory storage devices and other spintronics instruments, 
it is required, from one side, to have the possibility of keeping for long time 
a fixed spin alignment, and, from the other side, to be able to reverse the spin 
at any chosen moment of time. It is not a problem to block a spin for very long 
times using materials with large magnetic anisotropy. However, the same magnetic 
anisotropy does not allow for the realization of fast spin reversal. It would, 
of course, be possible to suppress the magnetic anisotropy by a strong external 
magnetic field, which, however, would need the use of very large fields, an order 
larger than the magnetic anisotropy. The optimal solution would be the use of 
minimally high magnetic fields for triggering spin reversal.

A method is advanced allowing for the initiation of spin reversal in nanomolecules 
and nanoclusters at any desired time. The idea of the method is to insert the 
sample inside a coil of electric circuit that creates a feedback field acting on 
spins and to arrange, at the required time, a resonance between the sample Zeeman 
frequency and the circuit natural frequency. This time-local resonance immediately 
triggers the start of spin reversal.

\newpage

\begin{center}
{\Large{\bf Figure Captions} }
\end{center}

\vskip 2cm
{\bf Figure 1}. Longitudinal spin polarization as a function of time for 
$\om=\om_0=10$, $\om_E=\om_1=0.01$, $\gm=1$, $A=1$, and different values of the 
switched on regulated field $b(\tau)=B$. The triggering resonance happens only 
for $B=A=1$.

\vskip 1cm
{\bf Figure 2}. Spin polarization as a function of time for the same parameters 
as in Fig. 1, but with the fixed triggering resonance condition $B=A=1$ and 
varying delay time $\tau$.   

\vskip 1cm
{\bf Figure 3}. Transverse spin component $x$ as a function of time for the same 
parameters as in Fig. 1, under the fixed triggering resonance condition $B=A=1$ 
and the delay time $\tau=10$. Figure $3a$ shows the time interval $[0,40]$ and $3b$, 
the time interval $[10,20]$. 

\vskip 1cm
{\bf Figure 4}. Time dependence of the feedback field $h$ for the parameters as 
in Fig. 3.  

\newpage

\begin{figure}[ht]
\centerline{
\hbox{ \includegraphics[width=10cm]{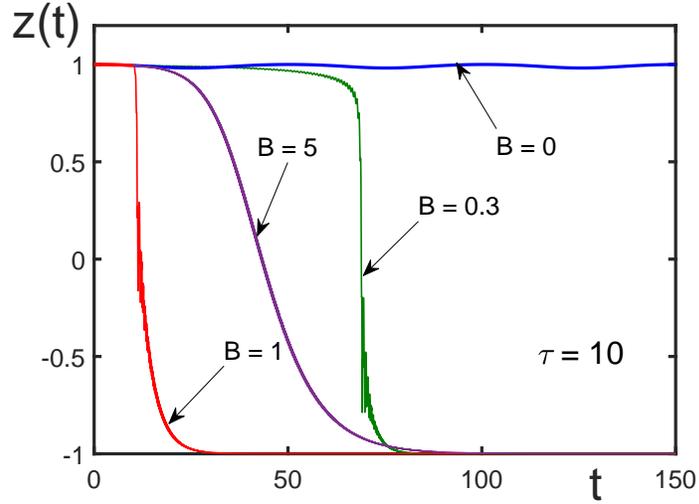}  } }
\caption{\small 
Longitudinal spin polarization as a function of time for 
$\om=\om_0=10$, $\om_E=\om_1=0.01$, $\gm=1$, $A=1$, and different values of the 
switched on regulated field $b(\tau)=B$. The triggering resonance happens only 
for $B=A=1$.
}
\label{fig:Fig.1}
\end{figure}

\begin{figure}[ht]
\centerline{
\hbox{ \includegraphics[width=10cm]{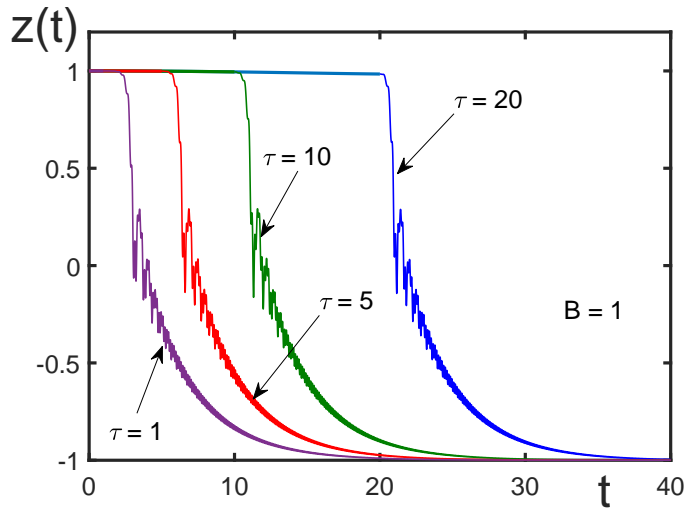}  } }
\caption{\small 
Spin polarization as a function of time for the same parameters 
as in Fig. 1, but with the fixed triggering resonance condition $B=A=1$ and 
varying delay time $\tau$. 
}
\label{fig:Fig.2}
\end{figure}

\begin{figure}[ht]
\centerline{
\hbox{ \includegraphics[width=7.5cm]{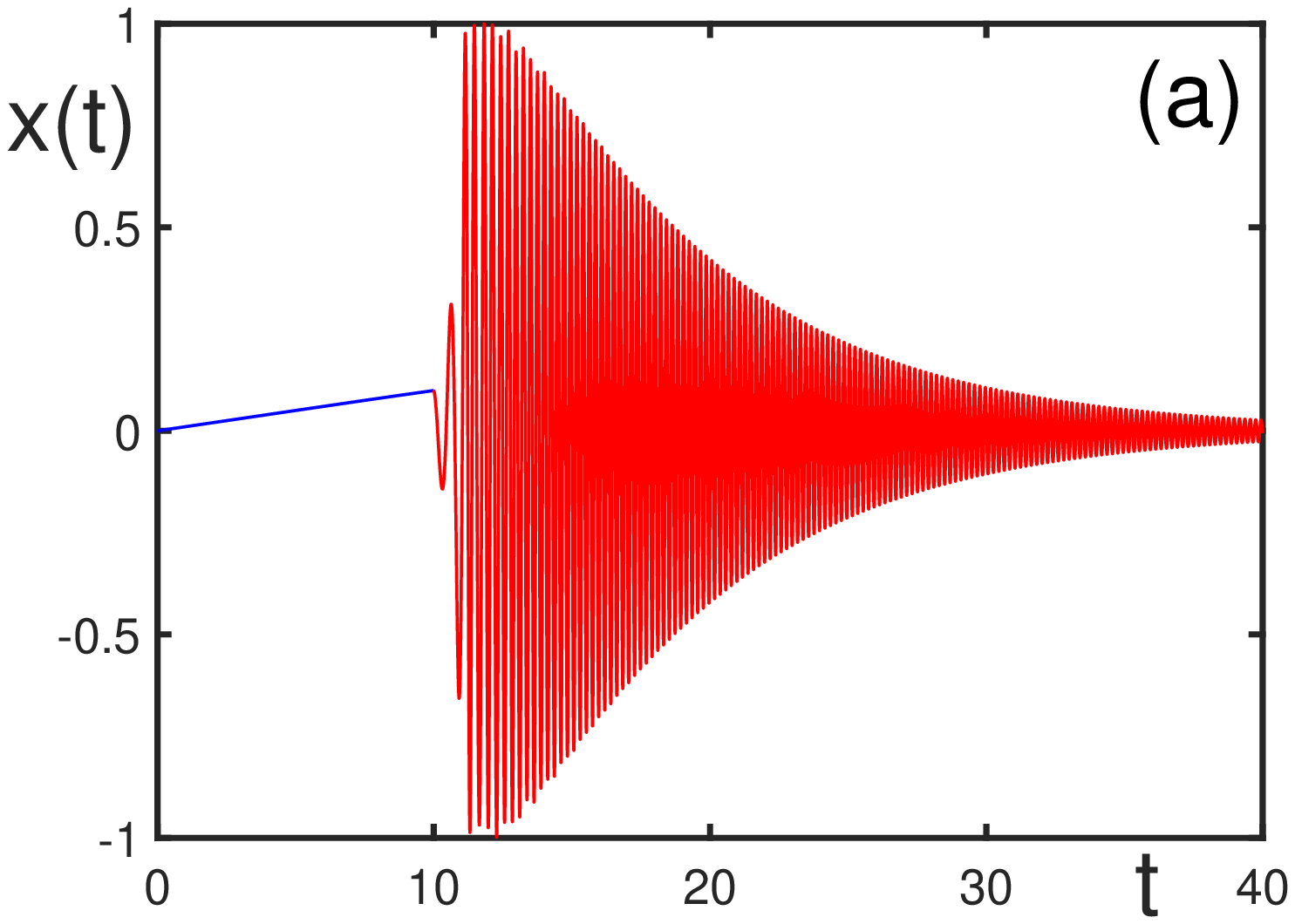} \hspace{1cm}
\includegraphics[width=7.5cm]{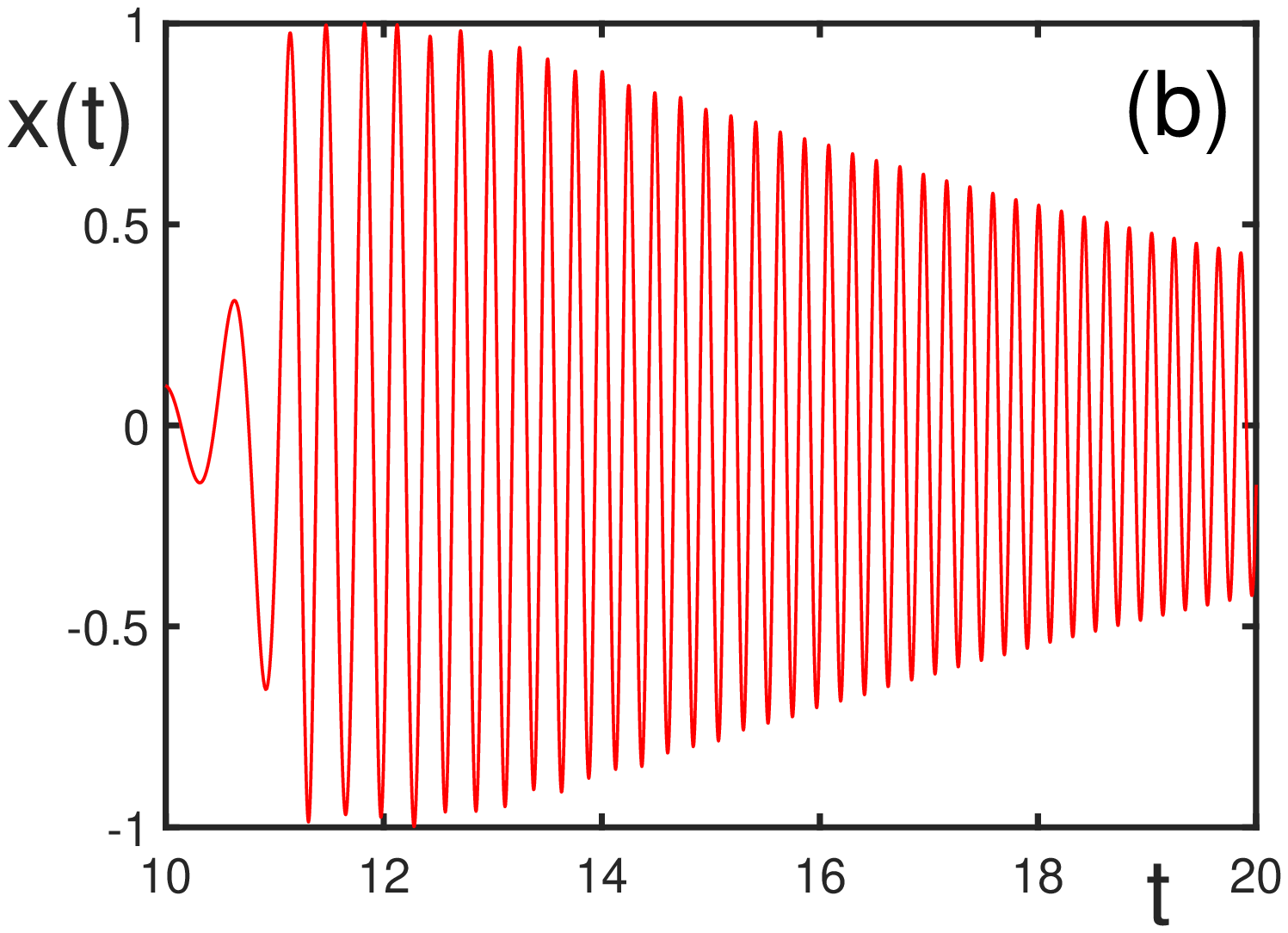}  } }
\caption{\small 
Transverse spin component $x$ as a function of time for the same 
parameters as in Fig. 1, under the fixed triggering resonance condition $B=A=1$ 
and the delay time $\tau=10$. Figure $3a$ shows the time interval $[0,40]$ and $3b$, 
the time interval $[10,20]$. 
}
\label{fig:Fig.3}
\end{figure}

\begin{figure}[ht]
\centerline{
\hbox{ \includegraphics[width=7.5cm]{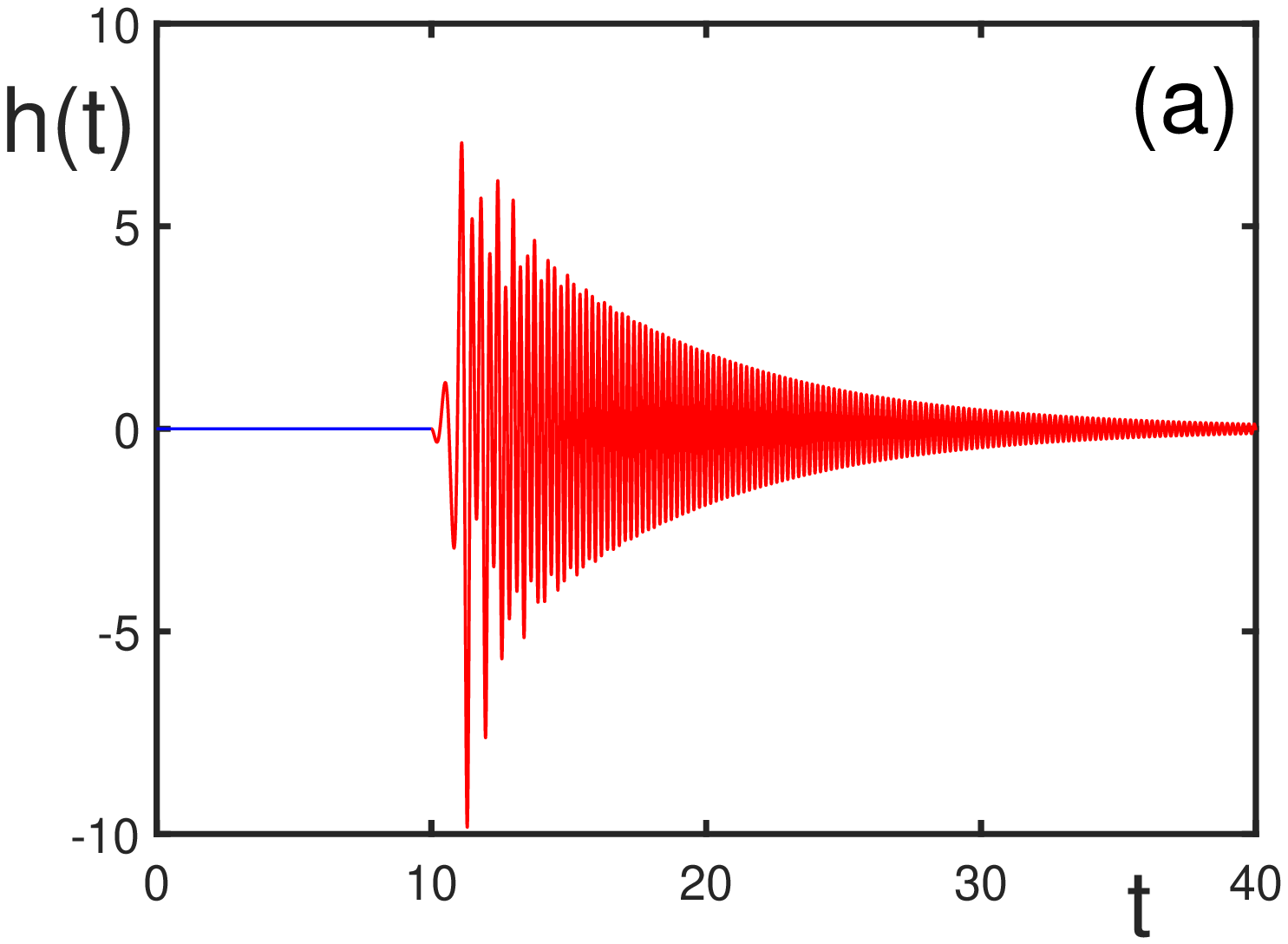} \hspace{1cm}
\includegraphics[width=7.5cm]{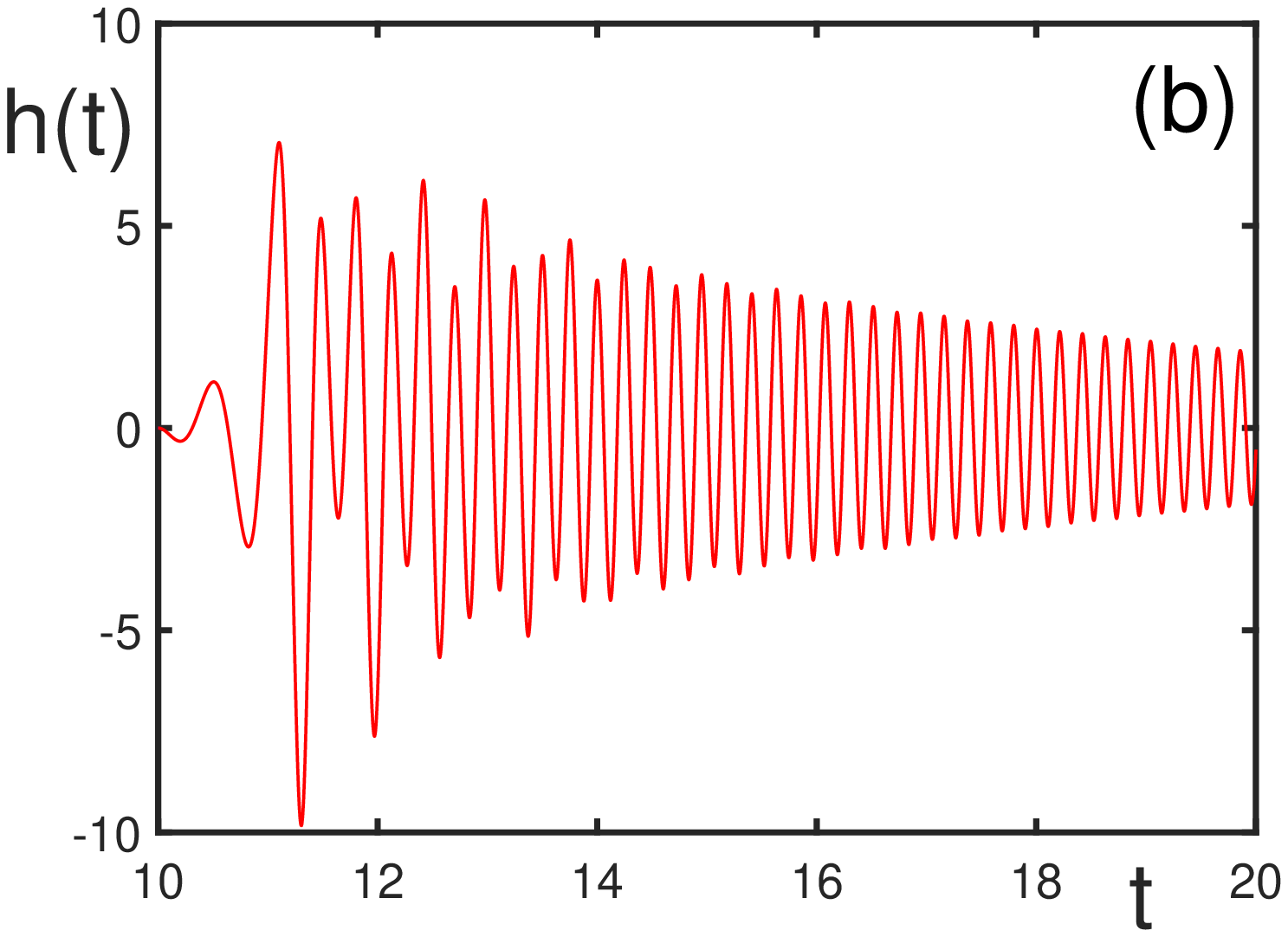}  } }
\caption{\small 
Time dependence of the feedback field $h$ for the parameters as in Fig. 3.  
}
\label{fig:Fig.4}
\end{figure}

\end{document}